\documentclass[conference]{IEEEtran}
\IEEEoverridecommandlockouts
\usepackage{cite}
\usepackage{amsmath,amssymb,amsfonts}
\usepackage{algorithmic}
\usepackage{graphicx}
\usepackage{textcomp}
\usepackage{xcolor}
\def\BibTeX{{\rm B\kern-.05em{\sc i\kern-.025em b}\kern-.08em
    T\kern-.1667em\lower.7ex\hbox{E}\kern-.125emX}}

\usepackage{acro}
\usepackage{array}
\usepackage{multirow}
\usepackage{caption}
\usepackage{float}
\usepackage{color,soul}
\usepackage{xurl}

\DeclareAcronym{ASiR}{short = ASiR , long = AirScale indoor radiohead}
\DeclareAcronym{BBU}{short = BBU ,  long = baseband unit}
\DeclareAcronym{RAN}{short = RAN , long = radio access network}
\DeclareAcronym{RAT}{short = RAT , long = radio access technology}
\DeclareAcronym{CN}{short = CN , long = core network}
\DeclareAcronym{3GPP}{short = 3GPP , long = Third Generation Partnership Project}
\DeclareAcronym{NR}{short = NR , long = new radio}
\DeclareAcronym{4G}{short = 4G , long = fourth generation}
\DeclareAcronym{5G}{short = 5G , long = fifth generation}
\DeclareAcronym{6G}{short = 6G , long = sixth generation}
\DeclareAcronym{3G}{short = 3G , long = third generation}
\DeclareAcronym{LTE}{short = LTE , long = long term evolution}
\DeclareAcronym{RF}{short = RF , long = radio frequency}
\DeclareAcronym{RRH}{short = RRH , long = remote radio heads}
\DeclareAcronym{pRRH}{short = pRRH , long = pico-RRH}
\DeclareAcronym{mRRH}{short = mRRH , long = micro-RRH}
\DeclareAcronym{PCI}{short = PCI , long = physical cell ID}
\DeclareAcronym{SSB}{short = SSB , long = synchronization signal block}
\DeclareAcronym{RSS}{short = RSS , long = root sum square}
\DeclareAcronym{LOS}{short = LOS , long = line of sight}
\DeclareAcronym{eMBB}{short = eMBB , long = enhanced mobile broadband}
\DeclareAcronym{mMTC}{short = mMTC , long = massive machine type communication}

\DeclareAcronym{EMF}{short = EMF , long = electromagnetic field}
\DeclareAcronym{SA}{short = SA , long = stand-alone}
\DeclareAcronym{NSA}{short = NSA , long = non-stand-alone}
\DeclareAcronym{NTN}{short = NTN , long = non-terrestrial network}
\DeclareAcronym{TN}{short = TN , long = terrestrial network}
\DeclareAcronym{ICNIRP}{short = ICNIRP , long = International Commission on Non-Ionizing Radiation Protection}
\DeclareAcronym{EU}{short = EU , long = European Union}
\DeclareAcronym{IEEE}{short = IEEE , long = Institute of Electrical and Electronics Engineers}
\DeclareAcronym{ICES}{short = ICES , long = International Committee on Electromagnetic Safety}
\DeclareAcronym{IEC}{short = IEC , long = International Electrotechnical Commission}
\DeclareAcronym{NATO}{short = NATO , long = North Atlantic Treaty Organization}
\DeclareAcronym{FCC}{short = FCC , long = Federal Communications Commission}
\DeclareAcronym{NCRP}{short = NCRP , long = National Council on Radiation Protection and Measurements}
\DeclareAcronym{WHO}{short = WHO , long = World Health Organization}
\DeclareAcronym{NDAC}{short = NDAC , long = Nokia Digital Automation Cloud}
\DeclareAcronym{TUK}{short = TUK, long = Technische Universit\"at Kaiserslautern}

\DeclareAcronym{PBCH}{short = PBCH , long = physical broadcast channel}
\DeclareAcronym{PDSCH}{short = PDSCH , long = physical downlink shared channel}
\DeclareAcronym{mMIMO}{short = mMIMO , long = massive multiple input multiple output}

\DeclareAcronym{UE}{short = UE , long = user equipment}

\DeclareAcronym{LEO}{short = LEO , long = low earth orbit}
\DeclareAcronym{MEO}{short = MEO , long = medium earth orbit}
\DeclareAcronym{GEO}{short = GEO , long = geostationary earth orbit}
\DeclareAcronym{NGSO}{short = NGSO , long = non-geostationary satellite orbit}
\DeclareAcronym{UAS}{short = UAS , long = unmanned aerial system }
\DeclareAcronym{HAPS}{short = HAPS , long = high altitude platform station}
\DeclareAcronym{SAN}{short = SAN ,  long = satellite access node}
\DeclareAcronym{NG-RAN}{short = NG-RAN , long = next generation radio access network}
\DeclareAcronym{IoT}{short = IoT , long = Internet of Things}
\DeclareAcronym{ITU}{short = ITU , long = International Telecommunication Union}

\DeclareAcronym{VSAT}{short = VSAT , long = very small aperture terminal }
\DeclareAcronym{ESIM}{short = ESIM , long = earth stations in motion}
\DeclareAcronym{EIRP}{short = EIRP , long = effective isotropic radiated power}
\DeclareAcronym{SAR}{short = SAR , long = specific absorption rate}

\DeclareAcronym{TDD}{short = TDD , long = time division duplex}
\DeclareAcronym{FDD}{short = FDD , long = frequency division duplex}
\DeclareAcronym{FR}{short = FR , long = frequency range}
\DeclareAcronym{FSS}{short = FSS , long = fixed satellite service}
\DeclareAcronym{MSS}{short = MSS , long = mobile satellite service}

\begin{document}

\title{Considerations on the EMF Exposure Relating to the Next Generation Non-Terrestrial Networks \\
\thanks{This work has been funded by the Federal Ministry of Education and Research of Germany (BMBF) (Projects numbers: 16KISK004 and 16KISK067). This is a preprint version, the full paper has been accepted by The IEEE International Conference on Mobile Ad-Hoc and Smart Systems (MASS), Toronto, Canada. September 2023. Please cite as: A. Fellan, A. Daurembekova and H. D. Schotten, “Considerations on the EMF Exposure Relating to the Next Generation Non-Terrestrial Networks,” in IEEE International Conference on Mobile Ad-Hoc and Smart Systems (MASS), 2023}
}

\author{\IEEEauthorblockN{Amina Fellan\textsuperscript{$*$}, Ainur Daurembekova\textsuperscript{$*$}, Hans D. Schotten\textsuperscript{$*$$\diamond$}}
\IEEEauthorblockA{\textsuperscript{$*$}\textit{Institute of Wireless Communication and Navigation,} \\ \textit{Rheinland-Pf\"{a}lzische Technische Universit\"{a}t Kaiserslautern-Landau (RPTU),} Kaiserslautern , Germany.\\
	\{fellan, daurembekova, schotten\}(at)eit.uni-kl.de}
\textsuperscript{$\diamond$}\textit{Intelligent Networks, German Research Center for Artificial Intelligence (DFKI),} Kaiserslautern, Germany.\\ Hans$\_$Dieter.Schotten(at)dfki.de}

\maketitle

\begin{abstract}
The emerging \ac{5G} and the upcoming \ac{6G} communication technologies introduce the use of space- and airborne networks in their architectures under the scope of \acp{NTN}. With this integration of satellite and aerial platform networks, better coverage, network flexibility and easier deployment can be achieved. Correspondingly, satellite broadband internet providers have launched an increasing number of small satellites operating in \ac{LEO}. These recent developments imply an increased \ac{EMF} exposure to humans and the environment. In this work, we provide a short overview of the state of consumer-grade satellite networks including broadband satellites and future \ac{NTN} services. We also consider the regulatory state governing their operation within the context of \ac{EMF} exposure. Finally, we highlight the aspects that are relevant to the assessment of \ac{EMF} exposure in relation to \acp{NTN}.    
\end{abstract}

\begin{IEEEkeywords}
non-terrestrial networks, mega-constellations, 6G, human EMF exposure, regulations, satellite networks 
\end{IEEEkeywords}

\section{Introduction}

Discussions about the next \acf{6G} of communications systems are well on course in both the academia and industry communities. Innovative solutions and disruptive technologies are being proposed to account for the shortcomings and challenges that faced previous generations. One of the challenges faced by current \acp{TN} is the difficulty and non-feasibility of providing global coverage to users in remote areas. The infrastructure deployment for \acp{TN} is often costly, rigid, and requires meticulous planning. Not to mention, in the case of natural disasters, how prone \acp{TN} are to breakdowns and how challenging it could be to restore their operation. 

\Acfp{NTN} are seen as one of the appealing solutions to complement \acp{TN} services and expand their coverage to remote areas around the globe. Recent advancements in the satellite and space industries lowered launch costs for satellite systems and advanced  the markets for consumer-service based deployments. Nowadays, the \ac{3GPP} is laying out the foundations for integrating \acp{NTN} in the next generation of communication networks. With \ac{6G}, two connectivity scenarios would be possible. Namely, direct connectivity between a satellite or an aerial platform and handset devices as well as indirect connectivity that exploit the satellites' connectivity in the backhaul \cite{kota}.

On the other hand, over the past decade, several companies announced revolutionary plans to provide high-speed internet access using constellations of a large number of satellites to serve user terminals around the globe. 
This increasing interest in bringing \acp{NTN} closer to end-users raises questions about how would such deployments affect the overall level of \acf{EMF} exposure experienced by humans and the environments. 

For \acp{TN}, organizations such as the \ac{WHO}, \ac{ICNIRP}, \ac{IEEE}, and \ac{ITU} are concerned with evaluating research and studies on human \ac{EMF} exposure \cite{who, ICNIRP2020, IEEEstd2019, itu_k9}. The outcomes of their evaluation is used to derive safe limits and guidelines to restrict the levels of \ac{EMF} exposure. In this work, we consider previous efforts done to study \ac{EMF} exposure related to satellite communications. 

This work is organized as follows. In Section \ref{satellite_networks}, we provide a summary on existing broadband internet satellite networks as well as the next generation \ac{NTN} networks as proposed by the \ac{3GPP}. Section \ref{measurements} explores considerations with respect to \ac{EMF} exposure from satellite and airborne networks and possible challenges related to \ac{EMF} measurements for such scenarios. The state of regulations concerning \ac{EMF} exposure within the scope of satellite communications is explored in Section \ref{regulations}. Finally, our conclusions are given in \ref{conclusions}.

\begin{table*}[h!]
\caption{Frequencies allocated to \acp{NTN} by the \ac{3GPP} \cite{3gpp.38.863}}
\begin{center}
\begin{tabular}{|c|c|c|c|c|c|}
\hline
\textbf{\acs{NR}} &\textbf{Satellite} &\multicolumn{2}{|c|}{\textbf{Frequency}}& \textbf{Duplex} & \textbf{Coexisting terrestrial } \\
\cline{3-4} 
\textbf{band} & \textbf{band} &\textbf{\textit{Uplink}}& \textbf{\textit{Downlink}}& \textbf{mode} & \textbf{\ac{NR} bands}\\
\hline
\textbf{n255} & L & 1626.5 MHz – 1660.5 MHz & 1525 MHz – 1559 MHz & \acs{FDD} &  n24*, n99* \\
\hline
\textbf{n256} & S & 1980 MHz – 2010 MHz & 2170 MHz – 2200 MHz & \acs{FDD} & n2, n25, n65, n66, n70 \\
\hline
\multicolumn{4}{l}{$^{\mathrm{*}}$US-specific.}
\end{tabular}
\label{tab_NTN_freq}
\end{center}
\end{table*}

\section{Non-Terrestrial Networks}\label{satellite_networks}
Interest in leveraging satellite and airborne networks for civilian applications has been gaining momentum over the recent years. Compared to \acp{TN}, satellite networks have the advantage of a larger coverage area which translates to better accessibility to remote and rural areas where the deployment of \acp{TN} can be challenging, not feasible, or not possible. \acp{NTN} could also ensure a fallback solution in natural disasters situations on the occasion that   \acp{TN}' services are disrupted. 
Within the scope of \ac{6G} networks, the introduction and integration of \acp{NTN} in the overall network architecture play an essential role in providing global network coverage and improving the network's resiliency. 
This will not be the first attempt to extend the coverage of \acp{TN} via satellite connectivity \cite{S-UMTS}, nor to provide mobile satellite services to users around the globe \cite{iridium}. However, the advent of the new low-cost space- and air-borne vehicles as well as the substantial development of communications technologies and markets makes the reintroduction of \acp{NTN} to support \acp{TN} quite attractive.

In this section we provide an overview of the state of the current and upcoming satellite networks, focusing particularly on those providing direct links of two-way satellite communications to user terminals and handheld devices. We first consider the \ac{3GPP}'s ongoing work on \acp{NTN}, followed by existing examples of mega-constellations of \ac{NGSO} satellite systems.

\subsection{New Radio - Non-Terrestrial Networks }

\begin{figure}[htbp]
    \centering
        \includegraphics[width=1\columnwidth]{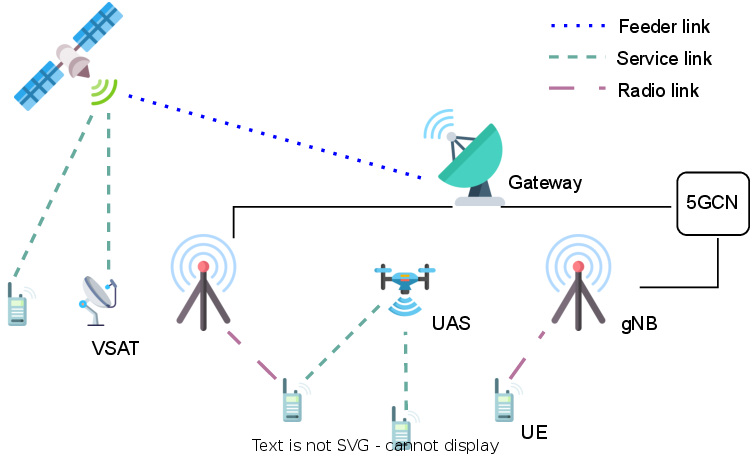}
        \caption{\ac{NTN} architecture according to the different scenarios considered by the \ac{3GPP}'s study in \cite{3gpp.38.811} } 
        \label{NTN_EMF}

\end{figure}

As per the \ac{3GPP}'s vision, the scope of \acp{NTN} includes spaceborne systems operating in the \ac{GEO}, \ac{NGSO} (including \acf{LEO} and \ac{MEO}), as well as airborne platforms such as \ac{HAPS} and \acp{UAS} \cite{zorzi2021}. The ongoing work on the \ac{NTN} standards can be broadly split into two classes, namely, \acf{NR}-\ac{NTN} and \ac{IoT}-\ac{NTN}. The former addresses \ac{eMBB} use cases, thus complementing and expanding the capacity of services provided by \acp{TN} over a larger coverage area. Whereas the latter is concerned with \ac{mMTC} applications, providing satellite connectivity to \ac{IoT} devices. 

The first studies considering the introduction of \acp{NTN} to the \acf{5G} \ac{NR} were initiated in 2017 by the \ac{3GPP} and appear in Rel-15 under TR 38.811 \cite{3gpp.38.811}. The \ac{3GPP} study considered use cases, propagation channel models, network architectures, deployment scenarios, and potential challenges associated with the integration of \acp{NTN} with the \ac{NR} interface. Figure \ref{NTN_EMF} illustrates some of the possible scenarios considered for \acp{NTN} based on \cite{3gpp.38.811}.

In Rel-16 under TR 32.321, the \ac{3GPP} resumed its investigation relating to the support of \ac{NR} protocols in \acp{NTN}, focusing mainly on satellite nodes but also considering other non-terrestrial platforms such as \acp{UAS} and \acp{HAPS} \cite{3gpp.38.821}. The technical report provided a refined proposal for \ac{NTN}-based \ac{NG-RAN} architectures, system- and link-level simulations for the physical layer, and considerations of issues relating to delay, Doppler shifts, tracking area and user mobility that are particular to \acp{NTN}.

Rel-17 builds upon the first studies carried out in \cite{3gpp.38.811} and \cite{3gpp.38.821} to define an initial set of specifications to support \acp{NTN} in \ac{NR}. Co-existence issues between \acp{NTN} and \acp{TN} channels as well as the \ac{RF} requirements for \acp{SAN} and \ac{NTN}-supporting \acp{UE} were the central point that the \ac{3GPP} considered in TR 38.863 \cite{3gpp.38.863}. Based on the \ac{ITU} radio regulations \cite{itu_rr}, the frequency bands n255 and n256 in the L- and the S-bands were designated by the \ac{3GPP} for \ac{NTN} operation in the United States and internationally, respectively. Details of the \ac{3GPP} approved allocated frequencies to \acp{NTN} are listed in Table \ref{tab_NTN_freq}. At present, \acf{FDD} channels are supported with plans to consider \ac{TDD} subsequently. For the operation at frequencies above 10 GHz, the satellite Ka-band is being considered \cite{3gpp.38.821}.

In the current Rel-18, the \ac{3GPP} is discussing in TR 38.882 the regulatory aspects and challenges arising from the verification of \acp{UE} location information within the coverage area of \acp{NTN} \cite{3gpp.38.882}. The definition and allocation of frequencies to the uplink and downlink in \ac{FR}2 is also currently under consideration. 

Despite the fact that the specifications and standardization of \acp{NTN} are at the moment still being defined by the \ac{3GPP}, the first initiatives to realize the \ac{3GPP}'s \ac{NTN} vision of a global hybrid mobile network are currently underway. For instance, the Omnispace Spark program has launched its first nanosatellite in 2022, part of a global \ac{NGSO} network and is foreseen to operate in the S-band, supporting the specifications of the \ac{3GPP}'s n256 band and targeting mainly \ac{IoT} applications \cite{omnispace}. 
Another milestone is the first \ac{UE} chipset supporting \ac{NR}-\ac{NTN}. It was announced in February 2023 by MediaTek \cite{mediatek}. The chipset is compliant with the \ac{3GPP}'s specifications for \ac{NR}-\ac{NTN} defined in Rel-17. It is capable of supporting two-way satellite communications and is seen as an enabler for \acp{NTN} services in future \ac{5G}  smartphones and satellite-enabled \acp{UE}.

\subsection{Satellite Networks}
Over the last decade, we witnessed a surge in the number of satellites being launched into space. With companies such as SpaceX, OneWeb, Telesat, and Blue Origin racing to kick-off their mega-constellations by sending hundreds of \ac{LEO} satellites to space \cite{megaconstellation}. The main goal of \ac{LEO} mega-constellations is to provide global broadband internet access to end-users over a \ac{FSS}. The architecture of mega-constellation satellite systems consists of three main components: the constellation of satellites in \ac{LEO}, a network of ground gateway stations, and user terminals; in the form of \acp{VSAT}. We consider here aspects of two of the currently largest constellations, namely, SpaceX and OneWeb.

In 2015, SpaceX announced its Starlink program with initial plans to deploy more than 4000 \ac{LEO} satellites in orbit. Starlink satellites are operated on circular orbits at an altitude of 550 km. As of the time of this writing, more than 3500 of Starlink's mini-satellites are in-orbit. The Ku and Ka satellite frequency bands are used for the satellite-to-user and satellite-to-ground links, respectively \cite{DELPORTILLO2019123}. The satellite and earth terminals support beamforming transmissions using phase-array antennas. Each satellite could support at least 8 beams. The peak \ac{EIRP} values for the user's downlink could reach 36.71 dBW \cite{DELPORTILLO2019123}.

OneWeb's constellation is planned to operate polar orbit planes at an altitude of 1200 km. A total number of 648 satellites are to be delivered in orbit to establish a global coverage of OneWeb's satellite network. Between the satellite and user terminals, the Ku band is used for communications while the Ka band is allocated for the satellite to gateway links. OneWeb satellites can support up to 16 beams. The maximum \ac{EIRP} for the user's downlink is around 34.6 dBW. 
Table \ref{starlink_oneweb} lists a comparison between characteristics of the two major \ac{NGSO} constellations currently in operation. 

\begin{table}[h!]
\caption{Summary of Starlink and OneWeb constellations operational characteristics}
\begin{center}
\begin{tabular}{ |c|c|c| } 
\hline
\textbf{Parameter}               &\textbf{Starlink} & \textbf{OneWeb} \\
\hline
\textbf{Altitude}                & 550 km           & 1200 km \\ 
\hline
\textbf{Satellite}               & 4408             & 648 \\ 
\textbf{constellation}           & (initial phase)  &  \\ 
\hline
\textbf{Number of user beams}    & $\geq$ 8             & 16  \\ 
\hline
\textbf{User-to-space}           &                   &  \\ 
\textbf{Uplink frequency}        & 14.0 – 14.5 GHz  &   14.0-14.5 GHz      \\ 
 \textbf{Downlink frequency}     & 10.7 – 12.7 GHz  &   10.7-12.7 GHz           \\ 
 \hline
 \textbf{Gateway-to-space}      &                  &   \\ 
  \textbf{Uplink frequency}     &  27.5 – 29.1 GHz & 27.5-30.0 GHz    \\ 
 \textbf{Downlink frequency}    & 17.8 – 18.6 GHz  & 17.8-19.3 GHz           \\ 
\hline
\end{tabular}
\label{starlink_oneweb}
\end{center}
\end{table}

Mega-constellation operators have been also extending their operations to include \ac{MSS} applications. SpaceX and T-Mobile announced in 2022 their plans to provide "Coverage Above and Beyond" services that support direct satellite connectivity to cellphone users over a \ac{LTE} interface \cite{spaceXTmobile}. The new service will be hosted on SpaceX's second generation \ac{NGSO} satellites and will support text messaging at the initial stage, with voice and data services to follow. The satellite-to-cellular links are planned to operate on \ac{LTE} band 25 frequencies, namely, 1910-1915 MHz and 1990-1995 MHz for the earth-to-space and space-to-earth links, respectively \cite{fcc2023}. 

Mega-constellation satellites are not the only nor the pioneer providers for \ac{MSS}. Commercial two-way communication for voice and data services was available by companies like Iridium, Inmarsat, Globalstar, and Thuraya ever since the late 1990s and early 2000s. For instance, the Iridium satellite constellation provides global voice and data services to mobile users via its 66 \ac{LEO} satellites \cite{iridium}. Its services are available to its subscribers on the L-band (from 1616-1626.5 MHz). In 2017, Iridium launched the second generation of its satellites, known as Iridium-NEXT \cite{iridium_web}.

\begin{figure}[htbp]
    \centering
        \includegraphics[width=\columnwidth]{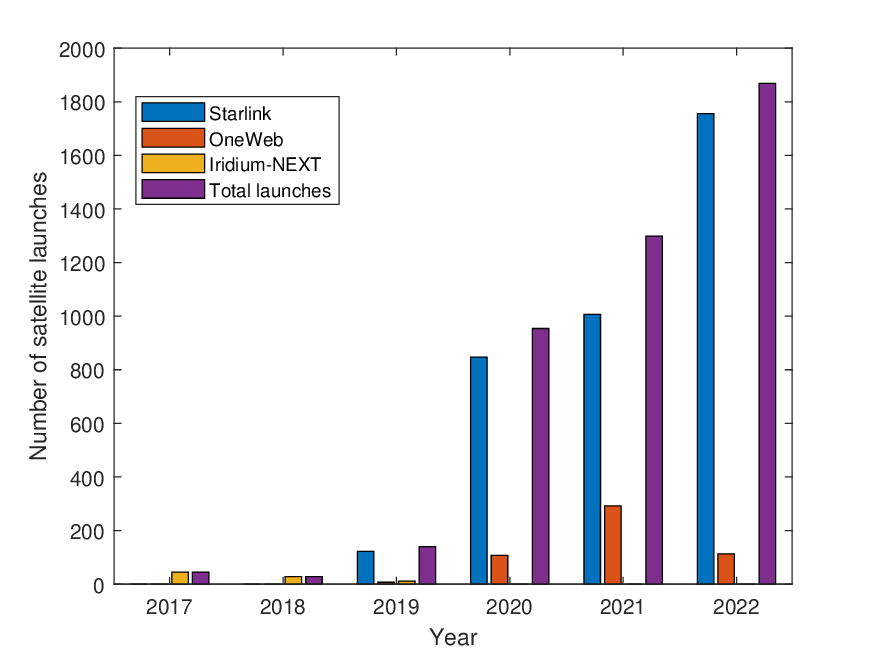}
        \caption{Number of satellite launches within the period of 2017-2022 for the three major satellite mobile communication providers. Based on data from \cite{gunter_space} } 
        \label{SatLaunch}

\end{figure}

Figure \ref{SatLaunch} depicts the number of satellites launches of the major mobile satellite communication providers over the last decade with a clear upward trend. In 2019, Iridium-NEXT has launched the required number of satellites to complete its planned constellation and provide a stable operation. Additional satellites could be launched in the future as spares to maintain the constellation. OneWeb and Starlink are still in the process of building their constellations.

\section{Measurements considerations and challenges}\label{measurements}

The recent growing interest in ubiquitous satellite communications with direct connectivity to \acp{UE} calls for more measurement and simulation studies on the exposure levels to users and the environment. Measurement studies on \ac{EMF} exposure near satellite earth stations are scarce and the emerging technologies to be implemented in both \acp{TN} and \acp{NTN} require more attention to their implications on the \ac{EMF} exposure levels. Table \ref{TN_vs_NTN} summarizes some of the aspects that might prove relevant when comparing between \ac{EMF} exposure assessment methods for \acp{TN} and \acp{NTN}. 

Hankin in \cite{Hankin1974}, supported by the US Environmental Protection Agency, provides one of the earliest and few measurement studies assessing the \ac{EMF} exposure levels from satellite communication systems. He first evaluated mathematically the \ac{EMF} exposure in terms of power density as a function of distance from the source, then he performed measurements in the vicinity of the sources to determine the actual exposure levels. However, the study considers satellite communication earth terminals transmitting at an \ac{EIRP} in the orders of megawatts, unlike the case for \acp{NTN} which will transmit at considerably lower powers. Moreover, such terminals are normally located away from the general population and are within sites that are restricted to public access. Besides, \ac{EMF} measurement equipment and satellite communication technologies have evolved significantly since the 1970s, when this study took place.  

More recent measurement studies were conducted in \cite{ijerph120505338} and  \cite{hussin2015_satelliteEMF}, where the authors evaluated the \ac{EMF} exposure using broadband \ac{EMF} meters in the vicinity of maritime satellite transmitters and satellite earth terminals, respectively. They later on compared it to the \ac{ICNIRP} guidelines levels for occupational and general public scenarios. 
In \cite{Novotny}, the authors consider the exposure from large-phased array antennas used to communicate with \ac{LEO} satellite systems and provide a device-based time-averaging method to estimate the \ac{EMF} exposure based on the time-averaged power density.

In this section we present a few considerations on \ac{EMF} exposure measurements for \acp{NTN} communications scenarios.

\subsection{Downlink: space-to-earth}
\begin{figure}[h]
    \centering
        \includegraphics[width=1\columnwidth]{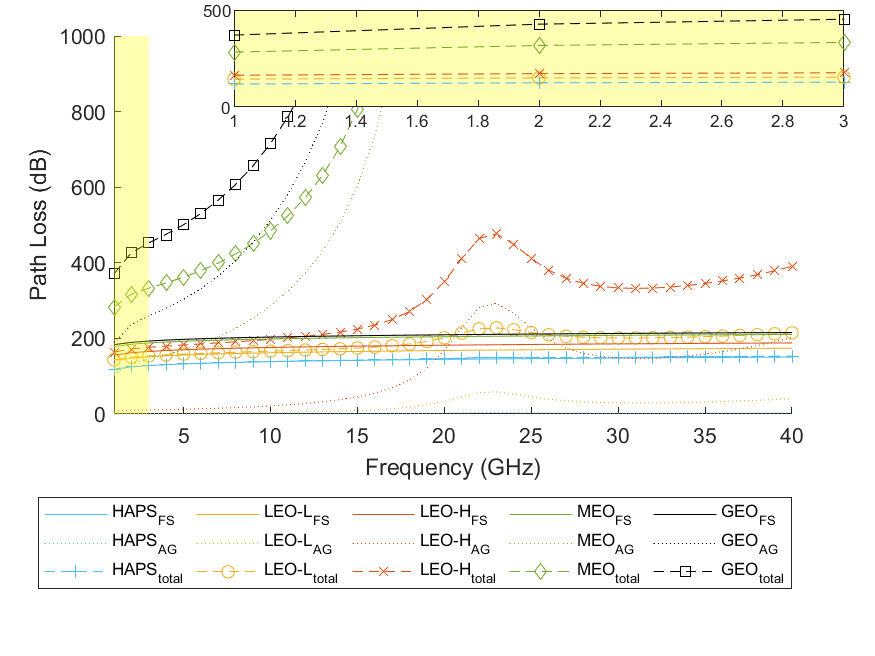}
        \caption{{Path loss due to free space (FS) and atmospheric gas (AG) conditions at a temperature of 20°C and  corresponding to the frequency range of satellite communications} } 
        \label{pathloss_sat}

\end{figure}

\textbf{Path loss}. By the time the signal arrives at the earth terminal, it would have experienced high losses due to several factors such as, free space path loss, atmospheric loss, and ionospheric and tropospheric scintillation losses. Free space path loss is dependant on the distance between the transmitter and receiver, which in the case of \acp{NTN} can be anywhere between 300 km for \ac{LEO} satellites and could reach up to 35,786 km for \ac{GEO} satellites. The free space path loss $L_{FS}$ is given by:

\begin{equation}
L_{FS}~(dB) = 20~log ( \frac{4~\pi~d~f}{c} ) 
\label{eq}
\end{equation}

where $d$ is the distance between the transmitter and receiver, $f$ is the frequency, and $c$ is the speed of light in vacuum. Atmospheric loss is the gaseous attenuation caused by oxygen and water vapor density \cite{itu_p676}. Whereas scintillation losses are caused by irregularities in the atmosphere \cite{itu_p618}. The overall path loss is dependant on the elevation angles of a satellite.
Figure \ref{pathloss_sat} depicts the path loss experienced at frequencies where satellite communications are operated, mainly from the L- to the Ka-bands. The highlighted and zoomed-in section corresponds to \ac{NR} bands n255 and n256 that are planned for \acp{NTN}' operation to support mobile satellite services. Five different elevations for \ac{NTN} space terminals are considered. Namely, 20 km for \ac{HAPS}, 300 km for the lower \ac{LEO} satellites altitude range, 1500 km for the upper \ac{LEO} satellites altitude range, 20,000 km for an average \ac{MEO} satellite altitude, and 35,786 km for \ac{GEO} satellites.

\subsection{Uplink: earth-to-space}
\textbf{Path loss} The uplink undergoes similar free space path loss to that experienced by the downlink. As a result, higher antenna gains are required at the user terminals transmitting from earth to space. 

\textbf{Terminal type}. Depending on the earth terminal type, the maximum transmitted power is defined accordingly. For instance, in \cite{3gpp.38.863} the \ac{3GPP} has assigned the \ac{UE} power class 3 to \acp{UE} for \ac{NTN} use-cases, which include handheld devices and allows a maximum transmit power of 26 dBm with a power tolerance level of +/- 2 dB for any transmission bandwidth within the channel bandwidth. \ac{UE} power class 3 is the default power class for \acp{TN} as well.
Other terminal types such as \acp{VSAT} and \acp{ESIM} are configured to transmit at higher powers. \acp{VSAT} and \acp{ESIM} would provide users indirect access to \ac{NTN} services. They are planned to be used for FR2 operation in \ac{NTN}. 

\textbf{Antenna type}. In contrast to \acp{UE}, \acp{VSAT} and \acp{ESIM} would use directional antennae such as parabolic antennae, i.e., the resulting \ac{EMF} exposure is concentrated in their antennae's main transmission lobe. Thus additional safety measures need to be taken into account when mounting such terminals to ensure the safety of users. 

\textbf{Service usage profiles}. For handheld \ac{UE} devices, a major component of the \ac{EMF} exposure occurs in the antenna's near-field. Adequate \ac{SAR} assessments methods are thus necessary in this case to determine the level of \ac{RF} power absorbed by the human body \cite{ICNIRP2020}. The usage profile (voice calls, texting, browsing, etc.) influences factors such as the proximity of the device to the body, the duration of usage, and consequently the intensity of \ac{EMF} exposure \cite{Varsier2015}.


\begin{table}[h!]
\caption{Comparison of some of the consideration on \acp{TN} vs \acp{NTN}}
\begin{center}
\begin{tabular}{ |c|c|c| } 
\hline
\textbf{Network}                &\textbf{\ac{TN}} & \textbf{\ac{NTN}} \\
\hline
\textbf{Antenna type}           & omni-directional          & directional \\
\hline
\textbf{Cell}                  & stationary                & stationary (GEO), \\
\textbf{coverage}              &                            & mobile (NGSO) \\
\hline
\textbf{Coverage }        & up to 30km               & up to 1000 km (NGSO), \\
\textbf{radius}            &                         & up to 3500 km (GEO) \\
\hline
\textbf{Frequency }        & FR1,                         & FR1, \\
\textbf{range (FR)}              & FR2                         & FR2* \\
\hline
\textbf{Operation }             & up to 7.125 GHz,          & 1-2 GHz \\
\textbf{frequencies}            &  24.25-71.0 GHz           & above 10 GHz* \\
\hline
\textbf{Duplex mode}            & TDD / FDD                  & FDD* \\
\hline
\textbf{Terminal}                & \ac{UE}                   & \ac{UE}, \\
\newline \textbf{type}   	       &                         & \acs{VSAT},  \\
                                    &                        & {ESIM} \\
\hline
\textbf{\ac{UE} power }             &  class 2,             & class 3 \\ 
\textbf{class}                      & class 3               &  \\
\hline

\multicolumn{3}{l}{$^{\mathrm{*}}$ NTN specifics are still underdevelopment by \ac{3GPP}.} \\

\end{tabular}
\label{TN_vs_NTN}
\end{center}
\end{table}

\section{Regulations and standardization}\label{regulations}

The recent advancements of wireless communications networks and their pervasiveness in our daily lives raised few calls from the public concerning the increased levels of \ac{EMF} exposure in the environment. Several organizations, such as the \ac{WHO}, \ac{ICNIRP}, \ac{IEEE}, \ac{ITU} and \ac{IEC}, evaluate continuously the research studying the influence of \ac{EMF} exposure on humans. They use the outcome of their evaluations to define and maintain the guidelines and recommendations with the goal to protect humans and the environment against any possible harmful effects caused by the exposure to radio-frequency electromagnetic fields \cite{ICNIRP2020, IEEEstd2019, itu_k100, emf_fellan}. 

The \ac{EMF} exposure limits set by the regulatory organizations target the overall maximum possible radiation at a given frequency and location. This maximum level should take into consideration the \ac{EMF} exposure due to all different sources and \acp{RAT} surrounding the point of evaluation. Also, it has to take into account characteristics of the transmissions present, for instance, whether they are continuous or pulsed \cite{ICNIRP2020}. In practice, such assessments are not necessarily straight-forward with regular measurement equipment as they only provide the instantaneous and average values of \ac{EMF} radiation at a given location. 

The \ac{ICNIRP} guidelines are the most widely adopted recommendations internationally. They define the reference levels for limiting the \ac{EMF} exposure from sources operating at frequencies up to 300 GHz \cite{ICNIRP2020}.
The reference levels are frequency dependant and can be evaluated in terms of the electric field, magnetic field, or the power flux density. Depending on the type of exposure in question, e.g., due to a handheld device versus due to a base station, they are defined for local and whole-body exposure.  Table \ref{icnirp_limits} lists the exposure limits as specified by the reference levels defined by the ICNIRP in \cite{ICNIRP2020}. For instance, the incident power density caused by a \ac{VSAT} terminal transmitting in the Ku-band, and taking into consideration any other sources transmitting in the same band, when measured in the far-field of its antenna (i.e., at a distance greater than $2~D~2~/~\lambda~~ (m)$; where $D$ is the largest dimension of the \ac{VSAT} antenna and $\lambda$ is the wavelength) should not exceed the limit of $10~W/m^2$ when averaged over a duration of 30 minutes and over the whole-body.   
On the other hand, for handheld devices, due to the fact that they are used in close proximity to the body and particularly to certain critical parts such as the head, reference levels concerned with local \ac{EMF} exposure apply in this case. Handheld devices must undergo rigorous \ac{EMF} exposure testing to determine the \ac{SAR} levels. As per the \ac{ICNIRP}, for the general public, the local \ac{SAR} limit for the head and trunk lies at $2~W/kg$ averaged over $10~g$ of cubic mass and for exposure intervals greater than or equal to 6 minutes \cite{ICNIRP2020}. 

\begin{table*}[h!]

\caption{ICNIRP reference levels for exposure to electromagnetic fields at frequencies relevant to satellite communication applications, as defined in \cite{ICNIRP2020}.}

\begin{center}

\begin{tabular}{cccccc}

\hline
\textbf{Exposure type}&\textbf{Frequency (MHz)}&{\textbf{Averaging time (min)}}&{\textbf{E-field strength (V/m)}}&{\textbf{H-field strength (A/m)}}&{\textbf{Power density (W/m\textsuperscript{2})}} \\
\hline

\multirow{2}{4em}{\textbf{Whole-body}}  & {400 - 2000 }   &  \multirow{2}{1em}{30}  &                       1.375\textit{f}\textsuperscript{1/2}    &  0.0037\textit{f}\textsuperscript{1/2}    & \textit{f}/200 \\
                                        &{2000 - 300000 }  &                        &   N/A                            &   N/A                                     &     10 \\ 
\hline

\multirow{3}{4em}{\textbf{Local}}       & { 400 - 2000 }   & \multirow{3}{1em}{6}    &   4.72\textit{f}\textsuperscript{0.43}    &  0.0123\textit{f}\textsuperscript{0.43}    & 0.058\textit{f}\textsuperscript{0.86} \\

                                        & { 2000 - 6000 }  &                       &   N/A        
                                        &   N/A                                     &     40 \\ 

                                        & { 6000 - 300000 }  &                       &   N/A        
                                        &   N/A                                     &     55/\textit{f}\textsuperscript{0.177} \\ 
\hline

\multicolumn{4}{l}

\end{tabular}
\label{icnirp_limits}

\end{center}
\end{table*}

The \ac{ITU} is the foremost authority regulating and organizing international access to the space spectrum \cite{itu_rr}. The \ac{ITU} also provides recommendations on measuring the \ac{EMF} radiation by broadcast stations and considerations regarding the placement of satellite earth stations in \cite{itu_bs1698}. The choice of the location for such stations needs to be carefully determined such that the near-field and transition zones are far from residential and industrial areas. Since satellite earth stations transmit to satellites at higher powers, it is very likely that the resulting \ac{EMF} field exceeds the reference levels. Thus, access to such facilities must be restricted against unauthorized personnel.    

With the introduction of \acp{NTN}, the existing guidelines and recommendations need to be re-assessed to take into account the increased number of satellites in space, and user terminals on earth and any resulting consequences on the \ac{EMF} levels. 

\section{Conclusions} \label{conclusions}

In this work, we presented a short overview of the state of the next generation \acp{NTN} and their position in future \ac{6G} networks. Given this growing interest in \acp{NTN}, questions on how to assess their possible contribution to the \ac{EMF} exposure on humans and the environment need to be taken into account. \acp{NTN} will bring direct connectivity of \acp{UE}, such as smartphones and tablets, to satellite and air-bone communication systems, which raises questions on the levels of \ac{EMF} exposures the users are subject to experience in such scenarios. We provided considerations regarding the implications of the spread of these technologies on \ac{EMF} exposure levels in the environment from the downlink and uplink perspectives. We observed there is a shortage of \ac{EMF} assessment studies concerned with \ac{NTN} scenarios. More measurement and simulation studies are required to better understand the possible \ac{EMF} exposure levels on the uplink and to ensure the compliance of such communications links with the recommended reference levels defined by the respective organizations.

\section*{Acknowledgment}

This work has been funded by the Federal Ministry of Education and Research of Germany (BMBF) partially under the projects “Open6GHub” (grant number 16KISK004) and 6G-TakeOff (grant number 16KISK067). The authors alone are responsible for the content of this paper.

\bibliography{References.bib} 
\bibliographystyle{ieeetr} 

\end{document}